# Statistical Assessment of Renewable Energy Generation in Optimizing Qatar Green Buildings*

Sara Zaina[a], Shima Sadaf [b], Ansaruddin Kunju[c], Mohammad Meraj[b], Devrim Unal [d], and Farid Touati [b]

[a] Department of Architecture and Urban Planning, Qatar University
[b] Department of Electrical Engineering, Qatar University
[c] Depatment of Chemical Engineering, Qatar University
[d] KINDI Center for Computing Research, Qatar University

sara.zaina@qu.edu.qa[a], s.sadaf@qu.edu.qa [b], aa1203886@student.qu.edu.qa[c], meraj@qu.edu.qa[b], dunal@qu.edu.qa[d], and touatif@qu.edu.qa[b]

*Abstract*—The residential electrical energy scheduling of solar Photovoltaics (PV) is an important research area of the modern green buildings. On the demand side, factors such as building load, and the renewable PV energy resources are integrated together as a nonlinear, indefinite and time varying complex system, which is very difficult to forecast and optimize. These energy sources are greatly depending on the climatic conditions. It further makes the residential building energy management complex. To address this problem, we present statistical models for the effective utilization of the renewables and reducing the burden on the distribution network. The effect of weather parameters such as temperature, dust in the air, humidity and solar irradiation on the green buildings energy infrastructure is taken into account. The details are analyzed and presented in the manuscript. The real time data is analyzed in the SPSS software tool. The presented results show that the statistical models are necessary for the controller to take action for the efficient and reliable integration of the renewables.

## I. INTRODUCTION

Doha's rapid urbanization is at the center of efforts to prepare for the FIFA 2022 World Cup. The rapid increase of urbanization, modernization, and population has led to a significant increase in the electricity demand within the country [1]. It is part of Qatar National Vision 2030 and the Qatar National Development Strategy II (NDS2 2017-2022), targeting the transformation of the country to reduce the per capita consumption rate and secure the demand for electric power to enhance sustainable development.

According to the International Energy Agency (IEA), by the year 2050, 11% of the global energy demand will be provided by the photovoltaics (PV) leading to reduction of 2.3Gt of carbon dioxide emissions per year. IEA estimated that existing buildings are responsible for more than 40% of the world's total primary energy consumption and for 24% of global carbon dioxide emissions [2]. Light of day in the Gulf Cooperation Council (GCC) region is about 4449h/year in which 70% of it is sunshine, results in receiving 6kWh/m$^2$/day [2]. Large-scale PV plants are located in desert land where full sunlight is available for the PV energy conversion. However, the high temperatures, very high relative humidity, dusty atmosphere and high solar radiation levels, may significantly decrease the production of PV energy which can vary from one place to the other within a geographical area. In order to understand the power consumption, PV generation and storage of the energy via batteries on weather conditions (climate change) vice versa. Thus, it is necessary to analyze the effect of one another and predict the accurate behavior of the load and the power distribution system [3].

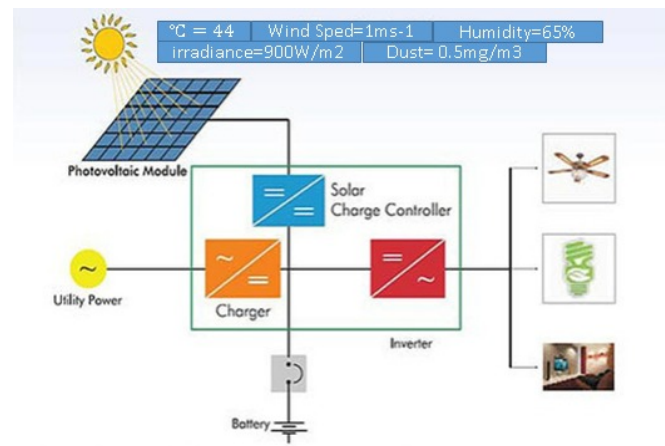

*Figure 1. Electricity Grid Residential with PV and Weather Conditions. [4]*

Atmosphere temperature is the key factor, which influences the energy demand. In countries with high average temperatures such as Qatar, most of the electricity consumption is due to cooling. It is well known that during hot days the power consumption is increased leading to more transmission and distribution losses and more peak tariffs. Although, the energy demand is dependent on the temperature, still the producers or researchers are not well informed with the temperature variations on energy consumption and energy storage. There is very little information about the relationship between the energy demand and the humidity [5]. The growing need of real-time energy management and automation systems resulted in green building energy management as one of the recent areas of attraction [6].

Distributed PV energy generation has been economically and environmentally beneficial solution for the residential and commercial consumers. In order to optimize the energy demand, the PV and the battery storage system has to be integrated in a full control manner [7]. These storage and the utility grid are connected in independently in islanded mode, depending on their PV – battery storage capacity and load consumption [8]- [9].

Besides, the relation among energy demand, PV generation and weather conditions is country dependent because, it depends on the geographical location, population, urbanism, style of the living and type of building code. So most of the published articles are country specific and are not generalized or universal models. The studies [10]- [11], developed error correction and estimation model for the energy consumption of Taiwan. In [12]- [13], authors studied about the variation of the weather data on electricity consumption in china and Hong Kong. In [14], authors found that, for small change in the climatic condition, there is huge change in the power consumption of the residential buildings in the United States. Saengngam [15] and Parkoom et al. [16] used multiple linear regression model to study the impact of the climate change on energy demand in Thailand. Wangpattarapong et al. studied the impact of climatic and economic factors on residential electricity consumption of the Bangkok Metropolis [17]. In [18], found that a long-run residential demand function for electricity and forecasting for the future load demand in Greece is sensitive to the actual income, price level, and weather conditions. They used ordinal regression analysis. Valor et al. [19] found a strong correlation between daily air temperature and electricity load in Spain. Blázquez et al. published a study on new empirical evidence using aggregated data on residential electricity demand in Spain [19]. The recent increase in use of Internet-of-Things (IoT) devices [20], [21] in the context of smart cities has also increased the energy demand.

In this article, the impact of weather conditions on solar energy generation and load consumption of Qatar green buildings are considered. The whole one year data November 20014 to October 2015 is considered for the analysis. The solar radiation is highest in Qatar than any other country in the world, which drives the researchers and the government to use sustainable green energy for the residential and commercial purposes. Simultaneously, the weather conditions like temperature, humidity, dust and raining all are affecting with wide variations in them on the energy demand of the Qatar. The accurate forecast of the weather data can help in optimizing the energy harvesting and can be freed from the carbon foot print in the region of Qatar. The effective estimation of the weather and amount of energy required can reduce the electrical power shortage and the risk of failures or the system collapse. Section II presents the geography of Qatar, per capita energy consumption, and Qatar national vision 2030 and legislation on the sustainable development. Statistical analysis for the data is carried out in section VI and their detailed discussions on the results. Finally, there are some concluding statements about the findings.

## II. Qatar's Geography, Climatic Conditions, and Energy Demand

Qatar is a peninsula in the east of Arabia, bordering the Persian Gulf and Saudi Arabia. It lies between latitudes 24°N and 27°N, and longitudes 50°E and 52°E. Majority of the country comprises of a low, barren plain, covered with sand dunes and salt flats [22].

Qatar is an arid country of harsh and fragile environment, high summer temperature (>40˚C), low summer humidity, low rainfall (annual average 71 mm) with high evaporation rate (annual average of 2,200 mm), and low nutrient availability in the soil. Alternatively, low temperature and high humidity during the winter season. The average high temperature is about 32.6°C and the average low temperature is around 21.4°C [22]. Wind acts as both friend and foe – as a cooling breeze, acknowledged during the summer season, or as biting cold amid winter storms [23].

Table 1 illustrate the maximum air temperature, humidity, irradiance, dust, and wind speed in the one-year data from November 20014 to October 2015, respectively. The maximum recorded temperature, humidity, irradiance, dust, and wind speed were 55°C, 92%, 1339W/m2, 3mg/m3, and 116km/h, respectively. The minimum recorded temperature, humidity, irradiance, dust, and wind speed were 8°C, 0%, 0W/m2, 0.02mg/m3, and 0km/h, respectively.

Table 1. Minimum and Maximum Weather Conditions. *(Source: Author's Creation)*

|  | Temperature (°C) | Relative Humidity (%) | Irradiance (W/m2) | Dust(mg/m3) | Wind Speed (km/h) |
|---|---|---|---|---|---|
| Minimum | 8.45 | .00 | .00 | .02 | .00 |
| Maximum | 54.90 | 92.43 | 1338.71 | 3.00 | 116.09 |

The 2016 census recorded the total population at 2,597,453 [24]. The population annual increase of 2016 is 7.3%. With increasing population and rapid urbanization, the electricity sector in Qatar has undergone a significant spurt in growth. This has led to a significant increase in the electricity demand within the country [1]. Qatar has been ranked the highest per capita electricity consuming country in the world. Thus, putting it on the top list of per-capita greenhouse gas emitting countries in the world. Alternatively, Doha's Government aims to reduce its carbon-dioxide footprint with the implementation of renewable energy and energy efficiency technologies [25].

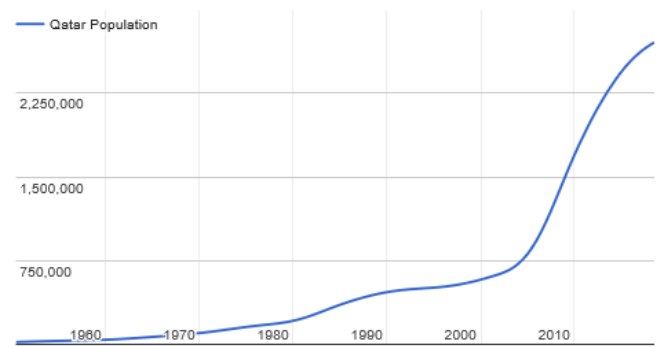

Figure 2. Qatar population growth rate against years from 1950-2018. *[26]*

Table 2. Average Electricity per capita Consumption. *[24]*

| Year | 2012 | 2013 | 2014 | 2015 | 2016 |
|---|---|---|---|---|---|
| Population | 1,836,676 | 2,045,239 | 2,235,431 | 2,421,055 | 2,597,453 |
| Population Annual Increase (%) | 4.4% | 11.4% | 9.3% | 8.3% | 7.3% |
| Total Energy Generation including all auxiliary consumption | 34,788 | 34,668 | 38,693 | 41,499 | 42,307 |
| Energy Taken into KM Network (Sent out) = Generation minus Auxiliary Consumption, kWh | 32,352 | 32,225 | 36,125 | 38,852 | 39,667 |
| Electricity Consumption, GWh (Excluding Bulk Industrial) | 20,387 | 20,121 | 22,216 | 24,491 | 25,108 |
| (A) Based on Production (Generation), (Includes Aux. Cons.) | 18,941 | 16,951 | 17,309 | 17,141 | 16,288 |
| (B) Based on Energy Taken into KM Network (Sent out) = Generation minus Aux. Cons. | 17,615 | 15,756 | 16,160 | 16,048 | 15,271 |
| (C) Based on Energy Taken into KM Network, Net of T&D Losses | 16,434 | 14,700 | 15,113 | 15,025 | 14,477 |
| (D) Based on Energy Taken into KM Network, Net of T&D Losses and Net of Bulk Industrial Consumption | 11,100 | 9,838 | 9,938 | 10,116 | 9,847 |

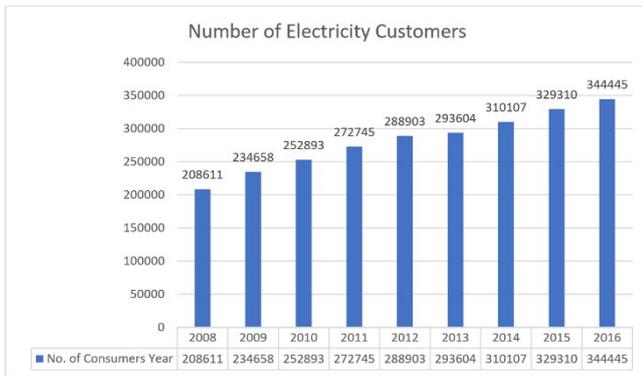

*Note that "Consumers" as used in this context is the number of customers registered with KAHRAMAA, not Qatar's population*

Figure 3. Number of Electricity Customers from 2008 to 2016. [1]- [24]

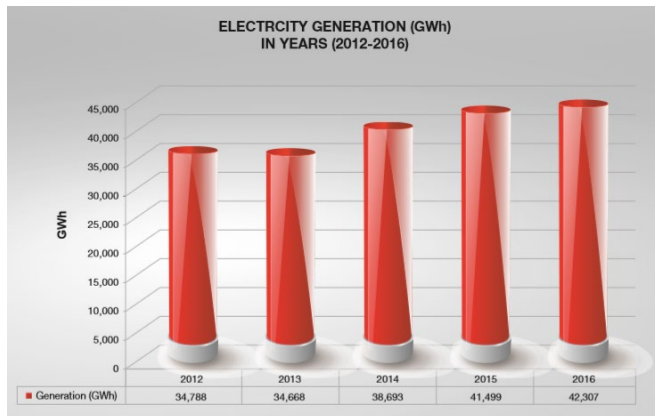

Figure 4. Electricity Generation (GWh) in Year 2012-2016. [24]

Figure 4 shows the sectoral consumption of electricity in 2016. It is evident that the highest electricity consuming sector is the domestic one followed by the industrial. These two sectors are known to be high consumers of electricity for the ventilation and air conditioning, which are very much dependent on the weather condition.

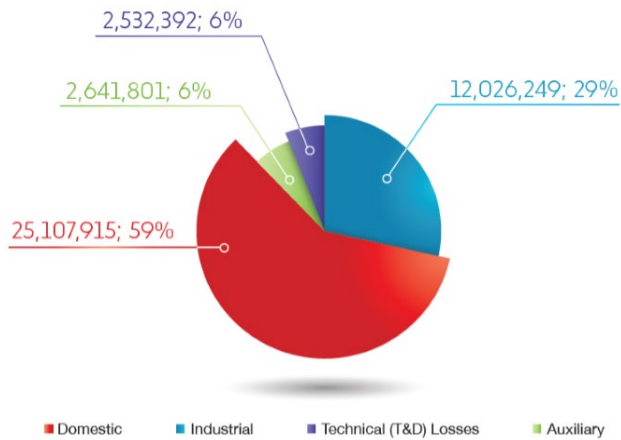

Figure 5. Sectoral Consumption in 2016, MWh. [24]

Qatar's government implements new sustainable means, national laws and strategies as a way to reduce the per capita consumption rate and secure the demand for electric power. Qatar has been committed to achieving sustainable development through the Qatar National Vision 2030 and the Qatar National Development Strategy II (NDS2 2017-2022) [1].

### III. THE RESEARCH DESIGN

This paper focuses on the impact of energy optimization on green buildings in Qatar. The study uses a combined research method of both qualitative and quantitative. Consequently, the overall research design for the collection of data for this study is structured into three phases.

(1) The review of the literature on the multi-disciplinary context providing a comprehensive set of data and providing secondary information on the geography of Qatar, weather conditions, per capita energy consumption, and Qatar national vision 2030 and legislation on the sustainable development and similar work published by [3]- [25]- [27].

(2) While we noted that some of the work on time-series prediction utilizes machine learning [28], this will be a subject of our future work.

(3) Touati, 2015 [27] collected the data for one year from November 2014 to October 2015. Two commercial PV modules were installed on the rooftop in a fixed position relative to the position of the sun. The position was situated to collect maximum solar radiations during the day. Then, all the sensors are installed and connected to the microcontroller for plotting P-V, I-V curves as well as collecting different parameters such as ambient and PV surface temperatures, irradiance, relative humidity, dust levels and wind speed. Also, a wireless system is developed with the help of LabVIEW and XBee (wireless module) for running the system accordingly [27].

(4) Data Cleaning
The pre-processing of data [29] was done considering the missing attributes and values, inconsistency, impossible and out of range values. Also, noisy data points were corrected by removing the errors, inaccurate values and outliers. In order to avoid the model mis-specification, biased parameter estimation and incorrect analysis results, outliers were eliminated.

(5) Statistical analysis using SPSS to explain the collection of data, including:
- Descriptive analysis (median, mode, mean, standard deviation, histograms)
- Box plots to identify outliers and distribution of means
- Scatter plots
- Bivariate Pearson correlation
- Multivariate linear regression
  *Note: assumption of 95% Confidence interval*

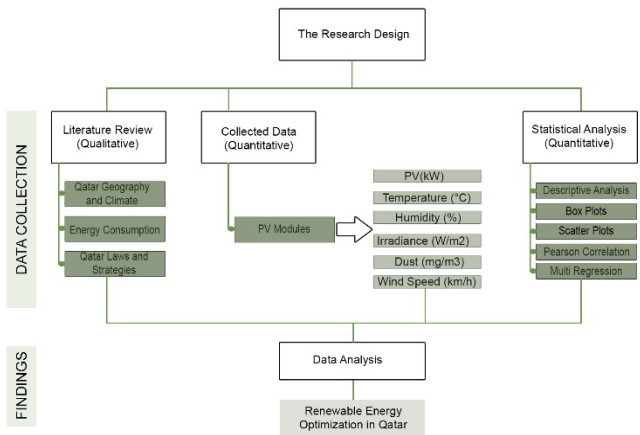

Figure 6. The Research Design Diagram.

Finally, the results enable to find the relationship between the weather conditions and energy demand in Qatar as demonstrated in the following section.

Variables that are analyzed from a statistical perspective include (1) weather conditions - temperature, humidity, irradiance, dust and wind speed; and (2) solar power generation (PV) and load.

## IV. DESCRIPTIVE STATISTICS AND DATA PROCESSING

### A. Descriptive Analysis

The analysis of the entire population of data has been selected to provide a more precise estimate of the data parameters such as the mean, median and standard deviation. Total 6830 observations were used to measure the central tendency. However, unusual values such as outliers, had a lesser impact on the median and mean due to the cleaning of data.

Table 3. Statistical Descriptive Analysis.

|  |  | Temperature (°C) | Relative Humidity (%) | Irradiance (W/m2) | Dust(mg/m3) | Wind Speed (km/h) | PV (kW) |
|---|---|---|---|---|---|---|---|
| N | Valid | 6830 | 6830 | 6830 | 6830 | 6830 | 6830 |
|  | Missing | 0 | 0 | 0 | 0 | 0 | 0 |
| Mean |  | 29.7239 | 45.3505 | 464.1698 | .5609 | 7.3360 | 9.61756 |
| Std. Error of Mean |  | .10395 | .22383 | 4.55886 | .00302 | .08415 | .090593 |
| Median |  | 29.4066 | 36.7640 | 446.9776 | .6437 | 6.2791 | 9.07304 |
| Mode |  | 24.96 | 32.64 | .00 | .82 | 3.08 | .000 |
| Std. Deviation |  | 8.59091 | 18.49826 | 376.76118 | .24987 | 6.95455 | 7.486917 |
| Variance |  | 73.804 | 342.186 | 141948.986 | .062 | 48.366 | 56.054 |
| Skewness |  | .001 | .864 | .142 | .301 | 7.596 | .284 |
| Std. Error of Skewness |  | .030 | .030 | .030 | .030 | .030 | .030 |
| Range |  | 46.45 | 92.43 | 1338.71 | 2.97 | 116.09 | 28.827 |
| Minimum |  | 8.45 | .00 | .00 | .02 | .00 | .000 |
| Maximum |  | 54.90 | 92.43 | 1338.71 | 3.00 | 116.09 | 28.827 |

Trimmed mean was calculated by discarding the top and bottom 5% of cases. Trimmed mean and mean were similar for the cleaned data as it suggests that there were few influential outliers.

The standard deviation for the relative humidity and irradiance were found to be high and indicates greater spread in the data, while the other variables such as temperature, solar power generation (PV), dust, and wind speed showed relatively reasonable spread of data. Large range value for some variables indicates greater dispersion in the data. However, some variables such as dust has small range value which indicates that there is less dispersion in the data.

### B. Distribution Analysis

The data distribution was assessed using both histogram and boxplots. The normality of data was examined by generating histogram over-laden with a normal curve and probability plot was not used as the entire population was used for analysis.

In the data set, distribution of the variables such as temperature, irradiance and solar power generation (PV) were found to be symmetric and normally distributed. Approximately, 68% of the values fall within one standard deviation of the mean, 95% of the values fall within two standard deviations, and 99.7% of the values fall within three standard deviations. Wind speed and relative humidity distributions were right skewed while the dust distribution was left skewed with majority of the data located on the higher side of the graph.

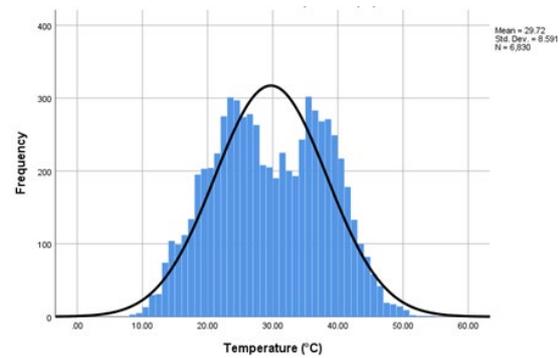
*(a) Frequency vs. Temperature*

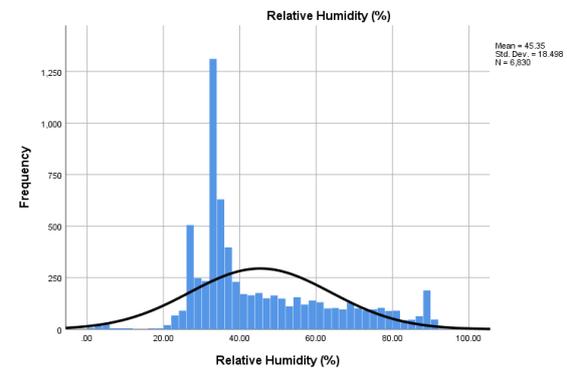
*(b) Frequency vs. Relative Humidity*

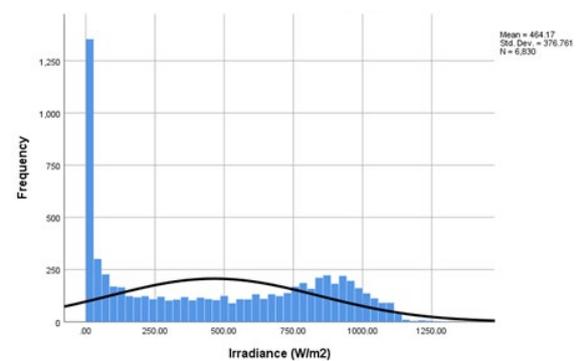
*(c) Frequency vs. Irradiance*

Figure 7. Histogram Distribution Diagrams.

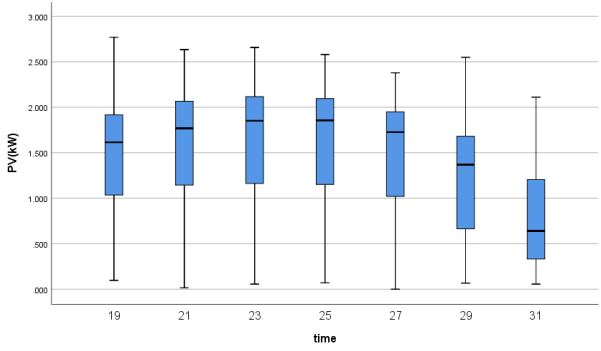

*(a) PV vs. Time*

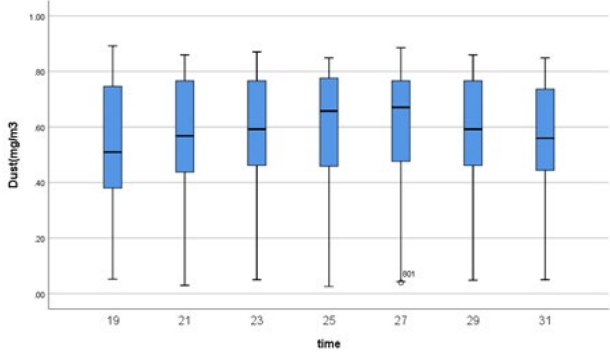

*(b) Dust vs. Time*

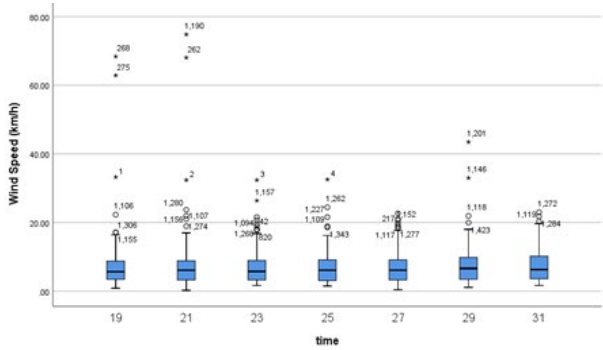

*(c) Wind Speed vs. Time*

Figure 8. Boxplots Distribution Diagrams.

## V. DETAILED STATISTICS ANALYSIS AND PREDICTIVE MODELS

### A. Homogeneity of Variances

After cleaning the data, the F ratio and the associated probability value (p-value) were calculated to test the null hypothesis for ANOVA that there is no significant difference among the groups (H0). It is noticed that the dependent variables such as temperature, irradiance and solar power generations are measured continuously with independence of observations and are approximately normally distributed with no significant outliers. The p-value associated with the F is smaller than 0.05 and nearly equal to zero and the null hypothesis is rejected. Hence, the alternative hypothesis is supported. When the null hypothesis is rejected, it is concluded that the means of all the groups are not equal. Moreover, the test of homogeneity of variances was done using Levine's test in SPSS and found that all the variables violate the homogeneity of variance assumption needed for ANOVA.

Table 4. Test of Homogeneity of Variances.

| | | Levene Statistic | df1 | df2 | Sig. |
|---|---|---|---|---|---|
| Temperature (°C) | Based on Mean | 13.088 | 24 | 6805 | .000 |
| | Based on Median | 11.565 | 24 | 6805 | .000 |
| | Based on Median and with adjusted df | 11.565 | 24 | 6006.280 | .000 |
| | Based on trimmed mean | 12.813 | 24 | 6805 | .000 |
| Irradiance (W/m2) | Based on Mean | 42.382 | 24 | 6805 | .000 |
| | Based on Median | 32.222 | 24 | 6805 | .000 |
| | Based on Median and with adjusted df | 32.222 | 24 | 4903.547 | .000 |
| | Based on trimmed mean | 39.381 | 24 | 6805 | .000 |
| PV (kW) | Based on Mean | 74.048 | 24 | 6805 | .000 |
| | Based on Median | 60.677 | 24 | 6805 | .000 |
| | Based on Median and with adjusted df | 60.677 | 24 | 5956.184 | .000 |
| | Based on trimmed mean | 74.539 | 24 | 6805 | .000 |

### B. Scatter Plots

Prior to investigating the relationship between irradiance and PV generation related to temperature, for humidity and wind speed a scatterplot was created - a graphical representation that includes both the variables and that provided a general illustration on the direction (positive association or negative association) of the relationship between the two variables.

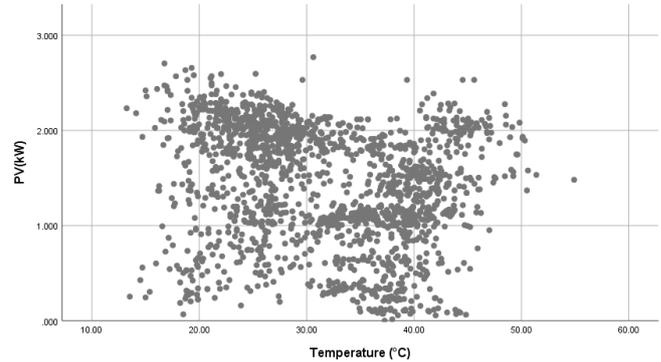

*(a) PV vs. Temperature*

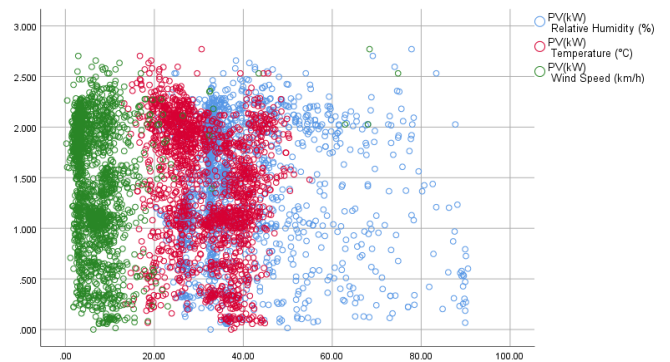

*(b) Relative Humidity, Temperature, and Wind Speed vs. Solar Power Generation*

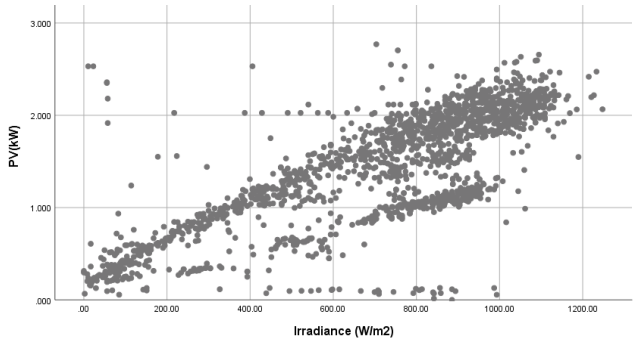

*(c) PV vs. Irradiance*

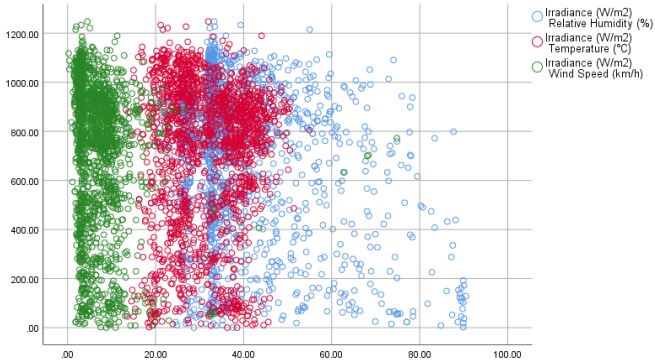

*(d) Relative Humidity, Temperature, and Wind Speed vs. Irradiance*

Figure 9. Scatter Plots.

## C. Bivariate Pearson Correlation

The strongest linear relationship occurs when the slope of the regression line is equal to 1. This means that when one variable increases by one, the other variable also increases by the same amount. The strength of the relationship between two variables is crucial and the interpretation of a scatterplot is too subjective. More precise evidence is needed, and this evidence is obtained by computing a coefficient that measures the strength of the relationship under investigation.

Table 5. Correlation Tables.

|  |  | PV (kW) | Temperature (°C) |
|---|---|---|---|
| PV (kW) | Pearson Correlation | 1 | .109** |
|  | Sig. (2-tailed) |  | .000 |
|  | N | 6830 | 6830 |
| Temperature (°C) | Pearson Correlation | .109** | 1 |
|  | Sig. (2-tailed) | .000 |  |
|  | N | 6830 | 6830 |

**. Correlation is significant at the 0.01 level (2-tailed).

|  |  | PV (kW) | Relative Humidity (%) |
|---|---|---|---|
| PV (kW) | Pearson Correlation | 1 | -.325** |
|  | Sig. (2-tailed) |  | .000 |
|  | N | 6830 | 6830 |
| Relative Humidity (%) | Pearson Correlation | -.325** | 1 |
|  | Sig. (2-tailed) | .000 |  |
|  | N | 6830 | 6830 |

**. Correlation is significant at the 0.01 level (2-tailed).

|  |  | PV (kW) | Wind Speed (km/h) |
|---|---|---|---|
| PV (kW) | Pearson Correlation | 1 | -.065** |
|  | Sig. (2-tailed) |  | .000 |
|  | N | 6830 | 6830 |
| Wind Speed (km/h) | Pearson Correlation | -.065** | 1 |
|  | Sig. (2-tailed) | .000 |  |
|  | N | 6830 | 6830 |

**. Correlation is significant at the 0.01 level (2-tailed).

|  |  | PV (kW) | Dust(mg/m3 |
|---|---|---|---|
| PV (kW) | Pearson Correlation | 1 | -.105** |
|  | Sig. (2-tailed) |  | .000 |
|  | N | 6830 | 6830 |
| Dust(mg/m3 | Pearson Correlation | -.105** | 1 |
|  | Sig. (2-tailed) | .000 |  |
|  | N | 6830 | 6830 |

**. Correlation is significant at the 0.01 level (2-tailed).

|  |  | PV(kW) | Irradiance (W/m2) |
|---|---|---|---|
| PV(kW) | Pearson Correlation | 1 | .756** |
|  | Sig. (2-tailed) |  | .000 |
|  | N | 1753 | 1753 |
| Irradiance (W/m2) | Pearson Correlation | .756** | 1 |
|  | Sig. (2-tailed) | .000 |  |
|  | N | 1753 | 1753 |

**. Correlation is significant at the 0.01 level (2-tailed).

|  |  | Irradiance (W/m2) | Temperature (°C) |
|---|---|---|---|
| Irradiance (W/m2) | Pearson Correlation | 1 | .273** |
|  | Sig. (2-tailed) |  | .000 |
|  | N | 6830 | 6830 |
| Temperature (°C) | Pearson Correlation | .273** | 1 |
|  | Sig. (2-tailed) | .000 |  |
|  | N | 6830 | 6830 |

**. Correlation is significant at the 0.01 level (2-tailed).

|  |  | Irradiance (W/m2) | Relative Humidity (%) |
|---|---|---|---|
| Irradiance (W/m2) | Pearson Correlation | 1 | -.409** |
|  | Sig. (2-tailed) |  | .000 |
|  | N | 6830 | 6830 |
| Relative Humidity (%) | Pearson Correlation | -.409** | 1 |
|  | Sig. (2-tailed) | .000 |  |
|  | N | 6830 | 6830 |

**. Correlation is significant at the 0.01 level (2-tailed).

|  |  | Irradiance (W/m2) | Wind Speed (km/h) |
|---|---|---|---|
| Irradiance (W/m2) | Pearson Correlation | 1 | -.061** |
|  | Sig. (2-tailed) |  | .000 |
|  | N | 6830 | 6830 |
| Wind Speed (km/h) | Pearson Correlation | -.061** | 1 |
|  | Sig. (2-tailed) | .000 |  |
|  | N | 6830 | 6830 |

**. Correlation is significant at the 0.01 level (2-tailed).

|  |  | Irradiance (W/m2) | Dust(mg/m3 |
|---|---|---|---|
| Irradiance (W/m2) | Pearson Correlation | 1 | .067** |
|  | Sig. (2-tailed) |  | .000 |
|  | N | 6830 | 6830 |
| Dust(mg/m3 | Pearson Correlation | .067** | 1 |
|  | Sig. (2-tailed) | .000 |  |
|  | N | 6830 | 6830 |

**. Correlation is significant at the 0.01 level (2-tailed).

|  |  | Load (kW) | Temperature (°C) |
|---|---|---|---|
| Load (kW) | Pearson Correlation | 1 | -.113** |
|  | Sig. (2-tailed) |  | .000 |
|  | N | 6830 | 6830 |
| Temperature (°C) | Pearson Correlation | -.113** | 1 |
|  | Sig. (2-tailed) | .000 |  |
|  | N | 6830 | 6830 |

**. Correlation is significant at the 0.01 level (2-tailed).

|  |  | Load (kW) | Relative Humidity (%) |
|---|---|---|---|
| Load (kW) | Pearson Correlation | 1 | -.025* |
|  | Sig. (2-tailed) |  | .037 |
|  | N | 6830 | 6830 |
| Relative Humidity (%) | Pearson Correlation | -.025* | 1 |
|  | Sig. (2-tailed) | .037 |  |
|  | N | 6830 | 6830 |

*. Correlation is significant at the 0.05 level (2-tailed).

|  |  | Load (kW) | Wind Speed (km/h) |
|---|---|---|---|
| Load (kW) | Pearson Correlation | 1 | -.060** |
|  | Sig. (2-tailed) |  | .000 |
|  | N | 6830 | 6830 |
| Wind Speed (km/h) | Pearson Correlation | -.060** | 1 |
|  | Sig. (2-tailed) | .000 |  |
|  | N | 6830 | 6830 |

**. Correlation is significant at the 0.01 level (2-tailed).

|  |  | Load (kW) | Dust(mg/m3 |
|---|---|---|---|
| Load (kW) | Pearson Correlation | 1 | -.169** |
|  | Sig. (2-tailed) |  | .000 |
|  | N | 6830 | 6830 |
| Dust(mg/m3 | Pearson Correlation | -.169** | 1 |
|  | Sig. (2-tailed) | .000 |  |
|  | N | 6830 | 6830 |

**. Correlation is significant at the 0.01 level (2-tailed).

As seen from the correlation tables, this is not a simple linear relation. A correlation coefficient measures the strength of that relationship. The Pearson correlation coefficient, r, can take a range of values from +1 to -1. A value of 0 indicates that there is no association between the two variables. A value greater than 0 indicates a positive association; that is, as the value of one variable increases, so does the value of the other variable such as solar power generation and irradiance with temperature.

An "r" of negative value indicates a negative linear relationship between variables such as relative humidity and wind speed with irradiance as well as solar power generation and shows statistical significance. The order of variables in this correlation is not important and provide evidence of association between the observed variables.

## D. Multivariate Linear Regression

Describing the relationship between two variables, correlations are not just sufficient. Other analysis should also be conducted to provide more information. Linear regression was used as step up after correlation for predicting the value of solar power generation and the load based on other independent variables such as temperature, relative humidity, wind speed, irradiance and dust. As there were more than two independent variables multiple regression analysis was used.

*1) Predictor: Solar power generation (PV)*

It was ensured that each predictor has a linear relation with the solar power generation which is the outcome variable and the variance of the errors are constant in the population. All the predictors such as temperature, relative humidity, irradiance, dust and wind speed were included to find the correlations with the solar power generation. The analysis of correlation showed that all predictor independent variables correlate linearly statistically insignificance with the outcome variable although there was substantial correlation among the predictors themselves. Table 6 indicates the four models that was selected to predict the outcome variable. From the model summary table (Table 6) the adjusted R square was found to increase from 0.711 to 0.730 when all the predictors were taken into account and thus it suggests that model 4 should be chosen.

Table 6. Model Summary for Solar Generation.

| Model | R | R Square | Adjusted R Square | Std. Error of the Estimate | R Square Change | F Change | df1 | df2 | Sig. F Change |
|---|---|---|---|---|---|---|---|---|---|
| 1 | .843a | .711 | .711 | 4.024890 | .711 | 16801.538 | 1 | 6828 | .000 |
| 2 | .853b | .727 | .727 | 3.913729 | .016 | 394.375 | 1 | 6827 | .000 |
| 3 | .854c | .730 | .730 | 3.891849 | .003 | 77.980 | 1 | 6826 | .000 |
| 4 | .855d | .731 | .730 | 3.886880 | .001 | 18.466 | 1 | 6825 | .000 |

a. Predictors: (Constant), Irradiance (W/m2)
b. Predictors: (Constant), Irradiance (W/m2), Temperature (°C)
c. Predictors: (Constant), Irradiance (W/m2), Temperature (°C), Relative Humidity (%)
d. Predictors: (Constant), Irradiance (W/m2), Temperature (°C), Relative Humidity (%), Wind Speed (km/h)
e. Dependent Variable: PV (kW)

Table 7. Coefficients for Solar Generation.

| Model | | Unstandardized Coefficients B | Std. Error | Standardized Coefficients Beta | t | Sig. |
|---|---|---|---|---|---|---|
| 1 | (Constant) | 1.840 | .077 | | 23.805 | .000 |
|   | Irradiance (W/m2) | .017 | .000 | .843 | 129.621 | .000 |
| 2 | (Constant) | 4.893 | .171 | | 28.591 | .000 |
|   | Irradiance (W/m2) | .017 | .000 | .879 | 133.659 | .000 |
|   | Temperature (°C) | -.114 | .006 | -.131 | -19.859 | .000 |
| 3 | (Constant) | 7.387 | .330 | | 22.404 | .000 |
|   | Irradiance (W/m2) | .017 | .000 | .859 | 124.557 | .000 |
|   | Temperature (°C) | -.147 | .007 | -.168 | -21.546 | .000 |
|   | Relative Humidity (%) | -.029 | .003 | -.073 | -8.831 | .000 |
| 4 | (Constant) | 7.468 | .330 | | 22.641 | .000 |
|   | Irradiance (W/m2) | .017 | .000 | .862 | 124.573 | .000 |
|   | Temperature (°C) | -.155 | .007 | -.178 | -21.909 | .000 |
|   | Relative Humidity (%) | -.031 | .003 | -.077 | -9.271 | .000 |
|   | Wind Speed (km/h) | .030 | .007 | .028 | 4.297 | .000 |

a. Dependent Variable: PV (kW)

The model 4 was adopted due to the incorporation of all the independent weather variables. The predicted value of the solar power generation can be computed as follows:
*Predicted solar power generation = 7.468 +*
*(0.017\*irradiance) + (-0.155\*temperature) +*
*(-0.031\*relative humidity) + (0.030\*wind speed)*

Table 8. Residuals Statistics for Solar Generation.

| | Minimum | Maximum | Mean | Std. Deviation | N |
|---|---|---|---|---|---|
| Predicted Value | -.58706 | 26.77897 | 9.61756 | 6.399605 | 6830 |
| Residual | -26.778973 | 25.397869 | .000000 | 3.885741 | 6830 |
| Std. Predicted Value | -1.595 | 2.682 | .000 | 1.000 | 6830 |
| Std. Residual | -6.890 | 6.534 | .000 | 1.000 | 6830 |

a. Dependent Variable: PV (kW)

Using the residual plots an analysis of the extent to which the predicted values differ from the actual values i.e. residuals were performed and the residuals were found to be having reasonable standard deviation of 3.89.

*2) Predictor: Load (KW)*
Similarly, it was ensured that each predictor has a linear relation with the Load which is our outcome variable and the variance of the errors are constant in the population. Predictors such as temperature, and relative humidity, were included and found to be linear.

The analysis of correlation showed that the predicted independent variables such as temperature and humidity correlate statistically, significantly with the outcome variable although there is substantial correlation among the predictors themselves. Only two models resulted in the outcome as the other variables were not statistically significant. The forward method was chosen and model 2 was selected to predict the outcome variable. From the model summary table (Table 9) the adjusted R square was found to increase from 0.013 to 0.026 when all the predictors were taken into account and thus it suggests that model 2 should be chosen.

Table 9. Model Summary for Load.

| Model | R | R Square | Adjusted R Square | Std. Error of the Estimate | R Square Change | F Change | df1 | df2 | Sig. F Change |
|---|---|---|---|---|---|---|---|---|---|
| 1 | .113a | .013 | .013 | 7.293392 | .013 | 88.049 | 1 | 6828 | .000 |
| 2 | .161b | .026 | .026 | 7.245513 | .013 | 91.538 | 1 | 6827 | .000 |

a. Predictors: (Constant), Temperature (°C)
b. Predictors: (Constant), Temperature (°C), Relative Humidity (%)
c. Dependent Variable: Load (kW)

Table 10. Coefficients for Load.

| Model | | Unstandardized Coefficients B | Std. Error | Standardized Coefficients Beta | t | Sig. |
|---|---|---|---|---|---|---|
| 1 | (Constant) | 10.927 | .318 | | 34.376 | .000 |
|   | Temperature (°C) | -.096 | .010 | -.113 | -9.383 | .000 |
| 2 | (Constant) | 15.614 | .583 | | 26.789 | .000 |
|   | Temperature (°C) | -.168 | .013 | -.197 | -13.278 | .000 |
|   | Relative Humidity (%) | -.056 | .006 | -.142 | -9.568 | .000 |

a. Dependent Variable: Load (kW)

The model 2 was adopted and the predicted value of the load can be computed as follows:
*Predicted Load = 15.6 + (-0.168\*temperature) +*
*(-0.056\*relative humidity)*

Table 11. Residuals Statistics for Load.

| | Minimum | Maximum | Mean | Std. Deviation | N |
|---|---|---|---|---|---|
| Predicted Value | 4.35411 | 12.51785 | 8.06133 | 1.178785 | 6830 |
| Residual | -11.381020 | 46.849194 | .000000 | 7.244452 | 6830 |
| Std. Predicted Value | -3.145 | 3.781 | .000 | 1.000 | 6830 |
| Std. Residual | -1.571 | 6.466 | .000 | 1.000 | 6830 |

a. Dependent Variable: Load (kW)

Using the residual plots an analysis of the extent to which the predicted values differ from the actual values i.e. residuals were performed and the residuals were found to be having reasonable standard deviation of 1.0 for the predicted value.

## VI. Conclusion

The weather data and the energy consumption for the residential building in Qatar were analyzed using the SPSS statistical tool. Results showed that weather parameters such as humidity, temperature, dust and irradiation are strongly correlated with the solar power generation. Whereas, the load demand correlates with temperature and humidity. However, the wind speed has no effect on the PV generation and the load demand (wind power generation unit is not considered for the residential application). It is also observed that the power consumption changes with climatic conditions. It's worth mentioning that the population growth rate for each year is significant and this was not considered in this research. This analysis is helpful in reducing the economic burden on the residential consumers and maximization of PV utilization. Nevertheless, this would foster reducing the carbon footprint in the state of Qatar by a better energy prediction and management. In the future we will investigate time-series prediction methods for more accurate and long-term predictions.

References

[1] "Sustainability Report," KAHRAMAA, Qatar, 2016.

[2] A. Aksakal and S. Rehman, "Global solar radiation in northeastern Saudi Arabia," *Renewable Energy*, pp. 461-472, 1999.

[3] F. Touati, M. A. Al-Hitmi and H. J. Bouchech, "Study of the Effects of Dust, Relative Humidity, and Temperature on Solar PV Performance in Doha: Comparison Between Monocrystalline and Amorphous PVS," *International Journal of Green Energy*, 2013.

[4] Indiamart, "Indiamart," Grid Connected Solar Power System, 2012. [Online]. Available: https://www.indiamart.com/proddetail/grid-connected-solar-power-system-7189644591.html. [Accessed 12 12 2018].

[5] "Qatar Statistics Authority," 9 May 2013. [Online]. Available: http://qsa.gov.qa.

[6] U. Hijawi, A. Gastli, R. Hamila, O. Ellabban and D. Unal, "Qatar green schools initiative: Energy management system with cost-efficient and lightweight networked IoT," in *2020 IEEE International Conference on Informatics, IoT, and Enabling Technologies (ICIoT)*, Doha, 2020.

[7] S.-J. Ahn, S.-R. Nam, J.-H. Choi and S.-I. Moon, "Power Scheduling of Distributed Generators for Economic and Stable Operation of a Microgrid," *IEEE Transactions on Smart Grid*, vol. 4, no. 1, pp. 398 - 405, 2013.

[8] K. Rahbar, C. C. Chai and R. Zhang, "Energy Cooperation Optimization in Microgrids With Renewable Energy Integration," *IEEE Transactions on Smart Grid*, vol. 9, no. 2, pp. 1482-1493, 2016.

[9] K. Rahbar, J. Xu and R. Zhang, "Real-Time Energy Storage Management for Renewable Integration in Microgrid: An Off-Line Optimization Approach," *IEEE Transactions on Smart Grid*, vol. 6, no. 1, pp. 124-134, 2015.

[10] P. Holtedahl and F. L. Joutz, "Residential electricity demand in Taiwan,," *Energy Economics*, pp. 201-224, 2004.

[11] Y.-K. Wu, G.-T. Ye and M. Shaaban, "Analysis of Impact of Integration of Large PV Generation Capacity and Optimization of PV Capacity: Case Studies in Taiwan," *IEEE Transactions on Industry Applications*, vol. 52, no. 6, pp. 4535-4548, 2016.

[12] W. L. Ping, "Study on the Relationship between Economic Development and Energy Consumption in Henan Province of China," in *2008 International Seminar on Business and Information Management*, China, 2008.

[13] W.Y.Fung, K.S.Lam, W.T.Hung, S.W.Pang and Y.L.Lee, "Impact of urban temperature on energy consumption of Hong Kong," *Energy*, vol. 31, no. 14, pp. 2623-2637, 2006.

[14] N. Lu, T. Taylor, W. Jiang, C. Jin, J. James Correia, L. R. Leung and P. Chun, "Climate Change Impacts on Residential and Commercial Loads in the Western U.S. Grid,"," *IEEE Transactions on Power Systems*, vol. 25, no. 1, pp. 480-488, 2009.

[15] N. Saengngam and U. Thonggumnead, "Predicting the medium-term electricity load demand of Thailand using the generalized estimating equation and the linear mixed effect model," in *2015 12th International Conference on Electrical Engineering/Electronics, Computer, Telecommunications and Information Technology (ECTI-CON)*, Thailand, 2015.

[16] S. Parkpoom and G. P. Harrison, "Analyzing the Impact of Climate Change on Future Electricity Demand in Thailand," *IEEE Transactions on Power Systems*, vol. 23, no. 3, pp. 1441-1448, 2008.

[17] K. Wangpattarapong, S. Maneewan, NiponKetjoy and W. Rakwichian, "The impacts of climatic and economic factors on residential electricity," *Energy and Buildings*, vol. 40, no. 8, pp. 1419-1425, 2008.

[18] D. Angelopoulos, J. Psarras and Y. Siskos, "Long-term electricity demand forecasting via ordinal regression analysis: The case of Greece," in *2017 IEEE Manchester PowerTech*, UK, 2017.

[19] E. Valora, V. Meneub and V. Caselles, "Daily Air Temperature and Electricity Load in Spain," *Journal of Applied Meteorology*, vol. 40, no. 8, pp. 1413-1421, 2001.

[20] M. Hammoudeh, G. Epiphaniou, S. Belguith, D. Unal, B. Adebisi, T. Baker, A. Kayes and P. Watters, "A service-oriented approach for sensing in the Internet of Things: intelligent transportation systems and privacy use cases," *IEEE Sensors*, vol. 1, no. 1, pp. 1-1, 2020.

[21] M. Zubair, D. Unal, A. Al-Ali and A. Shikfa, "Exploiting bluetooth vulnerabilities in e-health IoT devices," in *Proceedings of the 3rd International Conference on Future Networks and Distributed Systems*, Paris, 2019.

[22] Aquastat, "Geography, Climate and Population," Aquastat, Qatar, 2016.

[23] T. Dahl, Climate and Architecture, USA and Canada : Routledge, 2010 .

[24] "KAHRAMAA Annual Statistics Report," Qatar General Electricity & Water Corporation , Qatar, 2016 .

[25] A. Gastli, Y. Charabi, R. A. Alammari and A. M. Al-Ali, "Correlation Between Climate Data and Maximum Electricity Demand in Qatar," in *Research Gate*, Qatar, 2013.

[26] "worldometers," Qatar Population, 2018. [Online]. Available: http://www.worldometers.info/world-population/qatar-population/. [Accessed 6 December 2018].

[27] Touati, Farid; Al-Hitmi, M.A.; Chowdhury, Noor Alam; Hamad, Jehan Abu; Gonzales, Antonio J.R. San Pedro, "Investigation of solar PV performance under Doha weather using a customized measurement and monitoring system," *Renewable Energy,* 2015.

[28] M. Mudassir, S. Bennbaia, D. Unal and M. Hammoudeh, "Time-series forecasting of Bitcoin prices using high-dimensional features: a machine learning approach," *Neural Comput & Applic,* vol. 1, no. 1, pp. 1-1, 2020.

[29] D. Unal and M. U. Caglayan, "Spatio-temporal model checking of location and mobility related security policy specifications," *Turkish J Electr Eng Comput Sci,* vol. 21, pp. 144-173, 2013.